\documentclass{iopart}

\usepackage{graphicx}
\usepackage{subfigure}
\begin{document}

\article{Note}{Commissioning of a Geant4 based treatment plan simulation tool: linac model and dicom-rt interface}

\author{Iwan Cornelius$^{1}$, Brendan Hill$^{2}$, Nigel Middlebrook$^{2}$, Christopher Poole$^{1}$, Brad Oborn$^{3}$, Christian Langton$^{1}$}

\address{$^{1}$ Institute of Health and Biomedical Innovation, Faculty of Science and Technology, Queensland University of Technology, Brisbane, Australia.} 
\address{$^{2}$ Premion, Brisbane, Australia.} 
\address{$^{3}$ Department of Medical Physics, Illawarra Cancer Care Centre, Wollongong, Australia.}

\ead{iwan.cornelius@qut.edu.au}
\begin{abstract}
A Geant4 based simulation tool has been developed to perform Monte Carlo modelling of a 6 MV Varian$\textsuperscript{\textregistered}$ iX clinac. The computer aided design interface of Geant4 was used to accurately model the LINAC components, including the Millenium multi-leaf collimators (MLCs). The simulation tool was verified via simulation of standard commissioning dosimetry data acquired with an ionisation chamber in a water phantom. Verification of the MLC model was achieved by simulation of leaf leakage measurements performed using Gafchromic$\textsuperscript{\textregistered}$ film in a solid water phantom. An absolute dose calibration capability was added by including a virtual monitor chamber into the simulation. Furthermore, a DICOM-RT interface was integrated with the application to allow the simulation of treatment plans in radiotherapy. The ability of the simulation tool to accurately model leaf movements and doses at each control point was verified by simulation of a widely used intensity-modulated radiation therapy (IMRT) quality assurance (QA) technique, the chair test. 
\end{abstract}

\maketitle

\section{Introduction}

The Geant4 Monte Carlo toolkit was originally developed to support the high energy physics experiments of CERN (Allison \etal 2006); it has since enjoyed widespread usage in the medical physics community for many years (Foppiano \etal 2004, Paganetti 2004, Aso \etal 2007, Wroe \etal 2007, Oborn \etal 2009, Constantin \etal 2010). Geant4 possesses electromagnetic physics models that have been validated for materials and photon/electron energies relevant to radiotherapy (Poon and Verhaegen \etal 2005, Tinslay \etal 2007). Geant4 possesses many features that make it suitable for use in radiotherapy. It enables the simulation of complex geometries using combinatorial geometry, has support for voxelised geometries such as computed tomography (CT) data (Aso \etal 2007), and the ability to incorporate tesselated volumes generated by computer aided design (CAD) programs (Constantin \etal 2010). This capability is particularly useful in radiation detector development where complicated compositions and geometries can be modelled and sources of artefacts can be identified and mitigated (Othman \etal 2010). Time dependent geometries can also be modelled, which is applicable to modern radiotherapy modalities such as proton therapy (Paganetti 2004), tomotherapy, sliding window IMRT, volumetric modulated arc therapy, as well as tumour motion tracking technologies in these modalities. Lastly, Geant4 is able to model neutron production from photo-nuclear reactions, which is useful for out-of-field dosimetry studies relevant to higher energy photon beams. With this in mind, there is increasing interest in the use of Geant4 as a Monte Carlo tool for external beam radiotherapy (Jan \etal 2011, Grevillot \etal, Constantin \etal 2010, Foppiano \etal 2004). A recently developed multi-threaded version of Geant4 will enable porting of Geant4 applications to many core processing units, further placing MC in the reach of routine treatment plan verification studies (Dong \etal 2010). 

This note describes an accurate model of a Varian radiotherapy LINAC created using the Geant4 toolkit, integration with a DICOM-RT interface, and validation by comparison with experimental data. This tool will ultimately be used for routine independent verification of treatment plans and to support various research projects within the group including: gel dosimetry, plastic scintillator development, neutron dosimetry, and ultrasound based organ motion tracking. 

\section{Materials and methods}

\subsection{Geant4 LINAC simulation details}

\begin{figure}
\centering
\subfigure[]{
\includegraphics[width=0.4\textwidth]{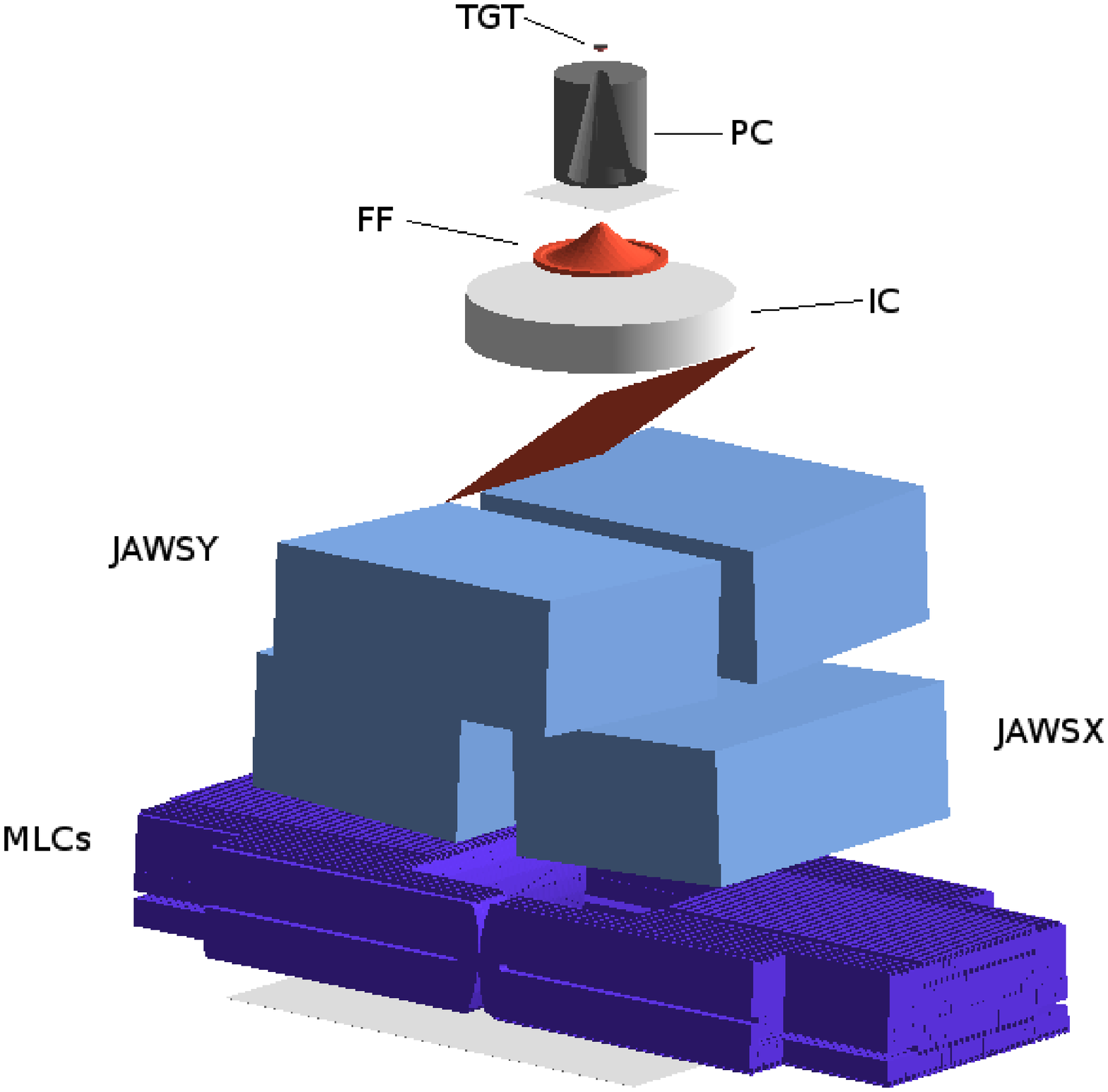}
}
\subfigure[]{
\includegraphics[width=0.4\textwidth]{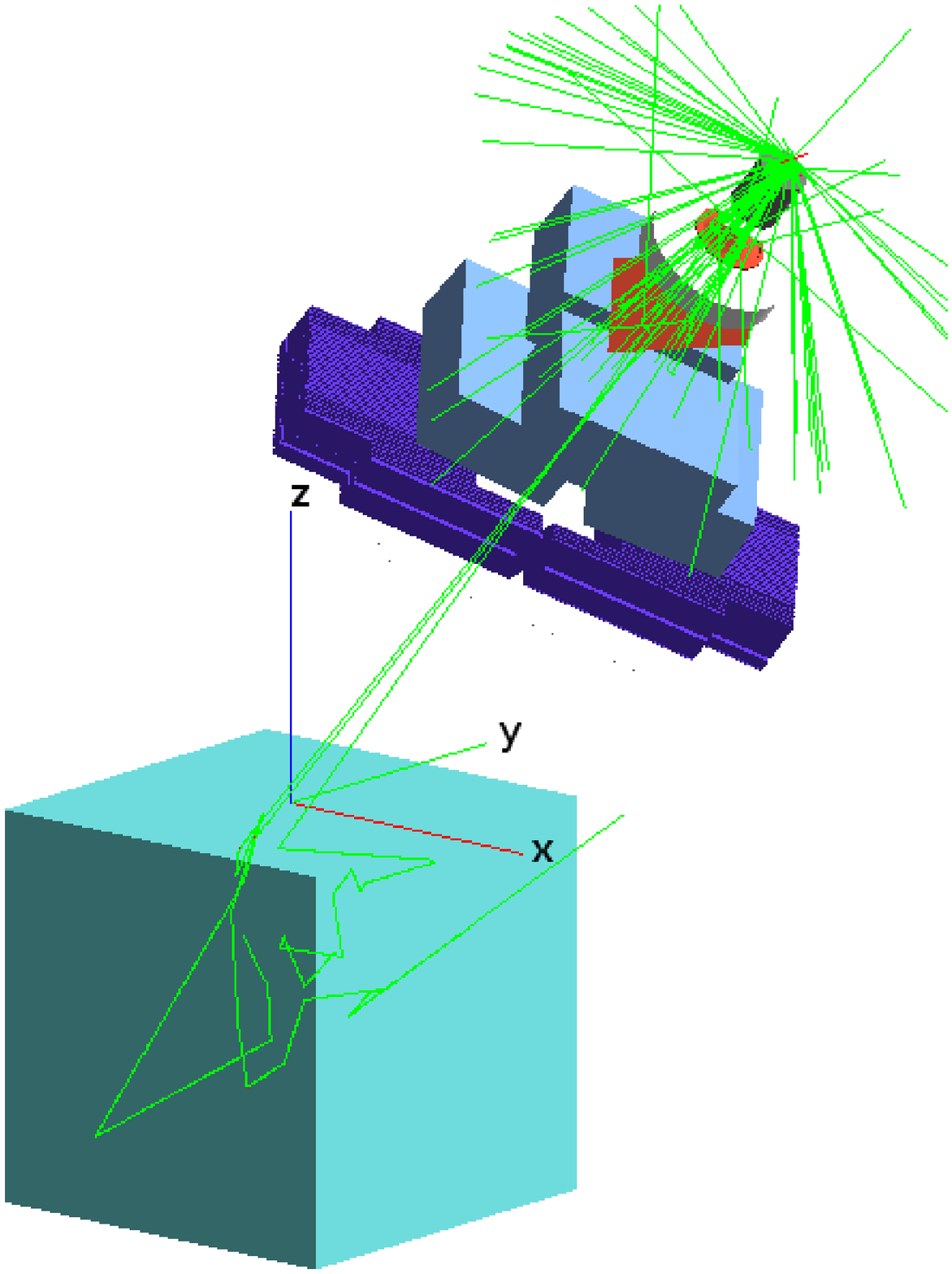}
}
\caption{(a) Geometry of the 6 MV Varian$\textsuperscript{\textregistered}$ iX clinac used in simulations including: target (TGT), primary collimator (PC), flattening filter (FF), ionisation chamber (IC), upper jaws in y direction (JAWSY), lower jaws in x direction (JAWSX), CAD modelled multileaf collimators (MLCs). nb. components are not to scale. (b) Positive gantry and collimator rotation in the IEC coordinate systems, room coordinate system is indicated. Also shown are a number of particle trajectories demonstrating interactions in the LINAC head and phantom readout geometry.}
\label{fig:geometry}
\end{figure}

The geometry of the 6 MV Varian$\textsuperscript{\textregistered}$ iX clinac model is illustrated in figure~\ref{fig:geometry} and is based on the vendor supplied documentation. Particular attention was paid to target, flattening filter, primary collimator, jaw, and MLC geometry and material composition. For the most part, Geant4's standard combinatorial geometry (Allison \etal 2006) classes were used to model the LINAC components. For complex IMRT and RapidArc treatments, contributions to the dose distribution from MLC leakage can be significant (Bush \etal 2008); an accurate model of MLCs is therefore essential. Each individual MLC leaf was modelled using Solidworks$\textsuperscript{\textregistered}$, including target and isocentre half-leaves, full leaves, and outboard leaves using the approach of Constantin \etal (2010). The leaf components were then exported individually as STEP files (an ISO compliant file format) before being converted to geometry description markup language (GDML) using the FASTrad$\textsuperscript{\textregistered}$ software. GDML is an XML extension used by Geant4 to allow definition of geometry without the need for hard coding. These GDML files were then loaded into the Geant4 application using the GDML parser. In order to create the MLC bank, each leaf was placed along the y-axis according to manufacturer specification, taking into account the interleaf gap and leaf divergence from the source. A number of user interface commands were developed to allow configuration of many aspects of the geometry without the need for recompilation between runs. This includes gantry angle, collimator angle, jaw and MLC positions, phantom composition and dimensions (solid water or water), phantom source-to-surface distance (SSD), and voxellisation of the readout geometry. 

The International Electrotechnical Commission (IEC 2002) has defined a standard for coordinate systems and positive rotation directions for gantry and collimator in radiotherapy. It provides a hierarchical approach, with each component coordinate system described relative to its mother coordinate system. Geant4 similarly uses a hierarchy of mother-daughter volumes in the construction of a simulation geometry, enabling straight-forward implementation of the IEC standard.

Standard parameterised electromagnetic physics models were used, taking into account the following processes: for photons, the photoelectric effect, compton scattering, rayleigh scattering, and pair production; for electrons, bremsstrahlung production, ionisation, and multiple scattering; and for positrons, multiple scattering, ionisation, and the annihilation process. To preclude the tracking of very low energy secondary particles, Geant4 uses the concept of range cuts; that is, if a secondary particle is produced with residual range less than the range cut, it is not tracked and assumed to deposit all energy at the point of generation; $100\, \mu m$ range cuts were used throughout the geometry. This corresponds to secondary electron energy thresholds of 84.7 keV in water, 352 keV in Tungsten, and 250 keV in Copper. 

Phase space files were used to reduce computation times. Using an approach similar to that of Bush \etal (2007), each simulation was executed in three phases. Phase I involves the simulation of a number of primary electrons incident on the target, the simulation of the subsequent electromagnetic cascade, and the scoring of photons at a plane below the ionisation chamber (scoring plane 1). This section of the geometry remains fixed during any treatment plan and need only be simulated once. The details of each particle crossing this scoring plane (position, direction cosines, energy, particle-type, particle weight) were recorded to a binary phase space file and the particle was no longer tracked. During phase II, particles were sampled from the first phase space file and transported through the jaws and MLCs, then recorded again at a second scoring plane below the MLCs (scoring plane 2). The region between scoring plane 1 and scoring plane 2 does not change between control points (beamlets) during an IMRT/RapidArc treatment. The first phase space file was recycled $N_{recyc}=25$ times to populate the second phase space file. Phase III of the simulation sampled the second phase space file and transported particles through the patient / phantom geometry; this phase space file was again recycled $N_{recyc}=25$ times. 

Uniform bremmstrahlung splitting (UBS) was implemented using the approach of Faddegon \etal (2008) with a splitting factor of $N_{ubs}=10$. A simple form of geometry biasing was also implemented via ``kill zones'': surfaces placed above the target and around the primary collimator so as to remove particles from the simulation that are unlikely to contribute to a response in the readout geometry. 

Values of $N_{recyc}$ and $N_{ubs}$ were optimised via calculation of simulation efficiency for a $10\times10\, cm^2$ square field using the methodology of Karakow and Walters (2006). 

Dose is scored in a voxellised water or patient geometry using the approach of Aso \etal (2007). The dose delivered to each voxel per primary electron is calculated for each control point of a treatment plan (or simulation run) along with an estimation of the standard error. 

All information needed to fully describe a radiotherapy treatment plan can be defined in a number of files using the DICOM-RT format. In order to facilitate a Monte Carlo simulation of treatment plans, a DICOM-RT interface is required. To this end, the VEGA library (Locke and Zavgorodni 2008) was incorporated into the Geant4 application. This library enables parsing of all DICOM-RT files as well as the exporting of an MC calculated dose distribution into a DICOM-RT DOSE file for subsequent importation back into the TPS. In our implementation, each beam (gantry angle, collimator angle) and control point (jaws, MLC settings, and number of MUs) is read from the TPS DICOM-RT PLAN file and translated into Geant4 user interface (macro) commands. These in turn modify the Geant4 simulation geometry between runs and simulate a pre-determined number of particles from the phase space file. 

Treatment plans generally give the absolute dose distribution within the phantom/patient, whereas the Monte Carlo calculation results in dose per primary particle. Analogous to the calibration of a LINAC monitor chamber, the virtual monitor chamber of a Monte Carlo simulation may be calibrated using the method outlined by Popescu \etal (2005). In this approach, the reference conditions of the LINAC are simulated, typically a $10\times10\, cm^2$ field at an SSD of 100 cm and calibration depth of $d_{max}$ (IAEA 2000). A simulation is executed to determine the dose per primary particle at the calibration depth, as well as the dose to the virtual monitor chamber per primary particle. The absolute dose distribution under general conditions can then be determined by:

\begin{equation}
D_{xyz,abs}=D_{xyz}\frac{(D^{forward}_{ch} + D^{back}_{ch^{(10\times10)}})}{(D^{forward}_{ch} + D^{back}_{ch})}\frac{D^{cal}_{xyz,abs}}{D^{cal}_{xyz}}U.
\end{equation}

Where $D_{xyz}$ is the normalised dose per incident particle, $D^{cal}_{xyz,abs}$ is the absolute dose per monitor unit as measured at the reference position during calibration of the linac, $D^{forward}_{ch}$ is the contribution to the monitor chamber dose per incident particle by the beam entering from above (phase I of the simulation), $D^{back}_{ch}$ is the contribution by the beam entering from the rear (phases II and III of the simulation), $D^{back}_{ch^{(10\times10)}}$ is this contribution under reference conditions, $D^{cal}_{xyz}$ is the simulated dose per primary particle under reference conditions, and finally $U$ is the number of monitor units delivered for a particular irradiation (or control point). 

\subsection{Commissioning}

According to the suggested protocol of Verhaegen and Seuntjens (2003) the first stage in the commissioning of a Monte Carlo linac model is to optimise primary electron beam parameters in order to obtain agreement with experimental results. These parameters are $E_e$ - the electron beam energy, and $\sigma_e$ - the standard deviation in the approximated gaussian fluence distribution of the primary electron beam that is normally incident on the target; ie., the spot-size. Dose distributions in a Scanditronix water phantom were measured using a Standard Imaging Exradin A16 chamber and Scanditronix IC13 chamber (for larger fields) as part of the commissioning procedure for the TPS. A subset of this was used for the tuning process; namely, a $1\times1 \, cm ^2$ and $10\times10 \, cm ^2$ square field irradiation of a water phantom. The dose distribution in a water phantom was simulated using a spatial resolution of $2\times2\times2\, mm ^3$. Initial values of $E_e=6\,MeV$ and $\sigma_e=1\,mm$ were based on manufacturer specifications. A comparison between simulated and experimental percentage depth dose (PDD) profiles, sensitive to beam energy, and a crossplane profile (sensitive to spot-size variations) at shallow depth ($d=15~mm$) was made. The spot-size and energy were tuned by nominal amounts until a $3\%/3\,mm$ gamma evaluation (Low \etal, 1998) criterion was achieved for 98\% of data points. 

Following primary beam tuning, a subset of the complete commissioning dataset was used for validation. Field-sizes relevant to IMRT and RapidArc treatments ranging from $1\times1~cm^2$ to $10\times10~cm^2$ were considered with the square fields defined by the X and Y jaws. Comparison was made using PDDs and cross-plane profiles at depths of $15~mm$, $50~mm$, $100~mm$, $200~mm$, and $300~mm$. Again a $3\%/3\,mm$ gamma evaluation criterion was used to quantify accuracy of the simulation and fine adjustments in the model were made where necessary.

In order to validate the CAD model of the multi-leaf collimator, a methodology similar to the approach of Heath and Seuntjens (2003) was used. Leaf leakage measurements were performed using EBT2 film placed at $50\, mm$ depth in a solid water phantom at SSD of $100 \, cm$. The phantom was irradiated by a $10\times 10\,cm^2$ jaw-defined field with MLCs fully closed with a collimator rotation of 90 degrees. Optical density changes in the EBT2 film were measured using an EPSON Perfection V700 scanner, 72 dpi resolution (spatial resolution of $0.35\,mm$) and 48 bit colour. Perspex frames were used to separate the film pieces from the scanner surface in order to avoid Newton's rings artefacts (Kairn \etal 2010). Analysis of the film pieces was conducted using ImageJ image processing software. Firstly the glass and frame were scanned to obtain a baseline image for the scanner, $P_{glass}(x,y)$. Next the film piece was placed on the frame and a scan acquired, $P_{film}(x,y)$. For both images, the red colour channel was isolated and converted to 32 bit. In order to calculate optical density the logarithm of the ratio of glass to film images was taken: $OD(x,y)= \log (\frac{P_{glass}(x,y)}{P_{film}(x,y)})$ (Hartmann \etal 2010). To calculate net optical density variation induced by radiation dose, the film was rescanned in this manner (placed at the same position on the frame as the pre-scanned film). Co-registration of images is performed (typically translation of the image by a few pixels) and the images are subtracted to calculate the net optical density image (Kairn \etal 2010). 

Calibration pieces of dimensions $3\times1$ cm$^2$ were used from the same sheet as the measurement pieces and were exposed to doses in the range 0-400 cGy using a $10\times10$ cm$^2$ square field at depth of $5 \, cm$ in a solid water phantom (at an SSD of $100 \, cm$). The pieces were scanned prior to irradiation and at $24$hrs after irradiation and net optical densities were calculated as above. A 2nd order polynomial fit to the calibration curve was made and used to convert net optical density to dose. To ensure that measured doses lay within the range of doses used for the calibration of film, two irradiations were conducted  (Heath and Seuntjens 2003). The first was approximately 50 times the number of MUs required to give an open field dose of 100 cGy, given that the dose under the leaves is approximately 2\% of the open field dose. Likewise, the dose under the abutted leaves is approximately 20\% of the open field dose, as such 5 times the MUs were considered for this measurement (Heath and Seuntjens, 2003). A simulation of the experiment was then performed.  

Validation of the LINAC model, DICOM-RT (PLAN) interface, and method of absolute dose calibration, was carried out by simulating a commonly used IMRT QA procedure known as the chair test (Van Esch \etal 2003). This test is commonly used to verify the correct functioning of the leaf motion controller, as well as correctness of treatment planning parameters such as transmission and dosimetric leaf separation. Similarly, it was used here to validate these aspects of the Geant4 simulation. A typical chair test plan was delivered to a MapCHECK$\textsuperscript{\textregistered}$ two dimensional detector array  at 5 cm depth in the MapPHAN$\textsuperscript{\textregistered}$ solid water phantom ($SSD = 95 ~ cm$), the plan was then simulated with the simulation tool, using the TPS generated DICOM-RT (PLAN) file for input. 

\section{Results and discussion}

\begin{figure}[here]
\centering
\subfigure[]{\includegraphics[totalheight=0.2\textwidth]{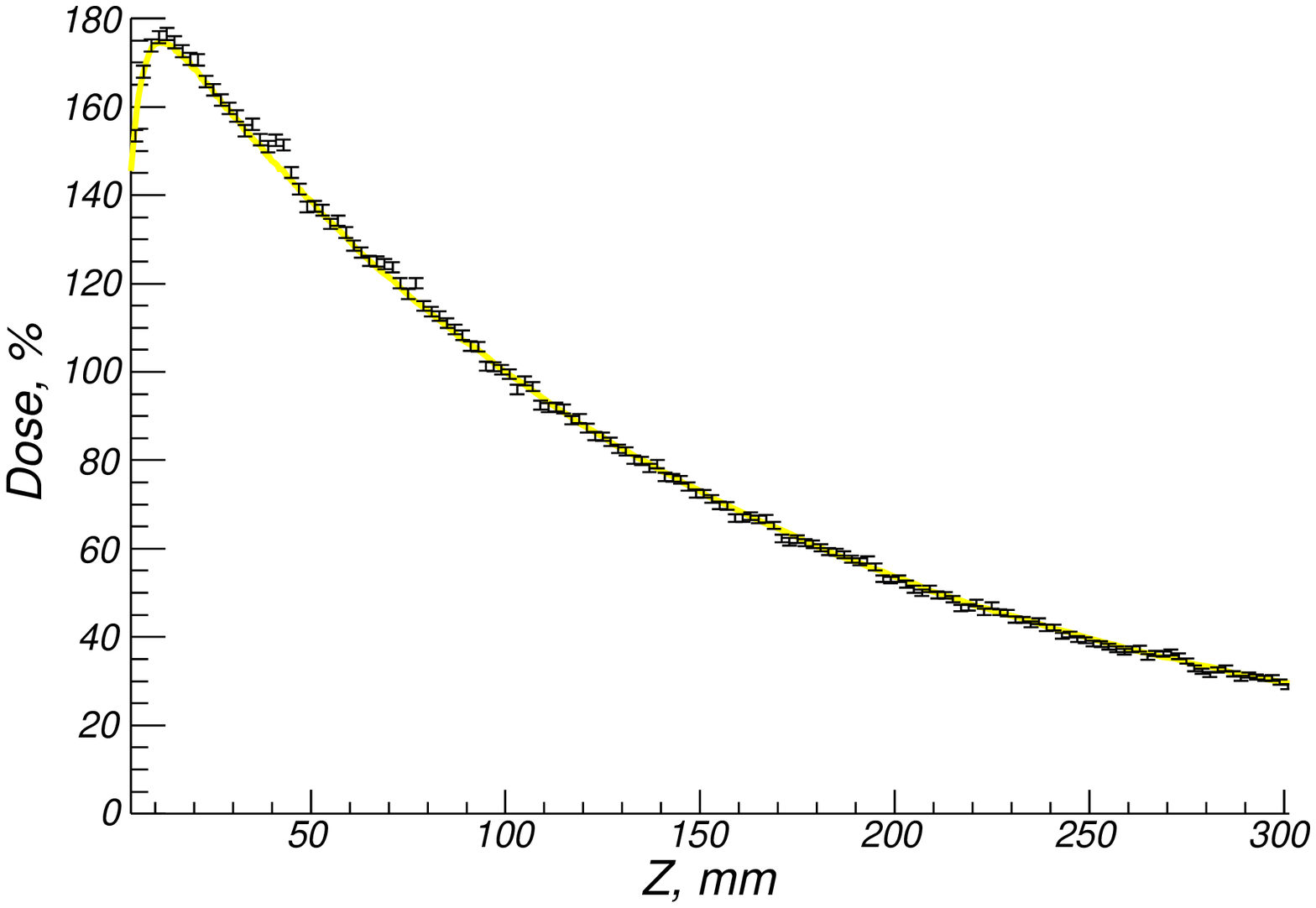}}
\subfigure[]{\includegraphics[totalheight=0.2\textwidth]{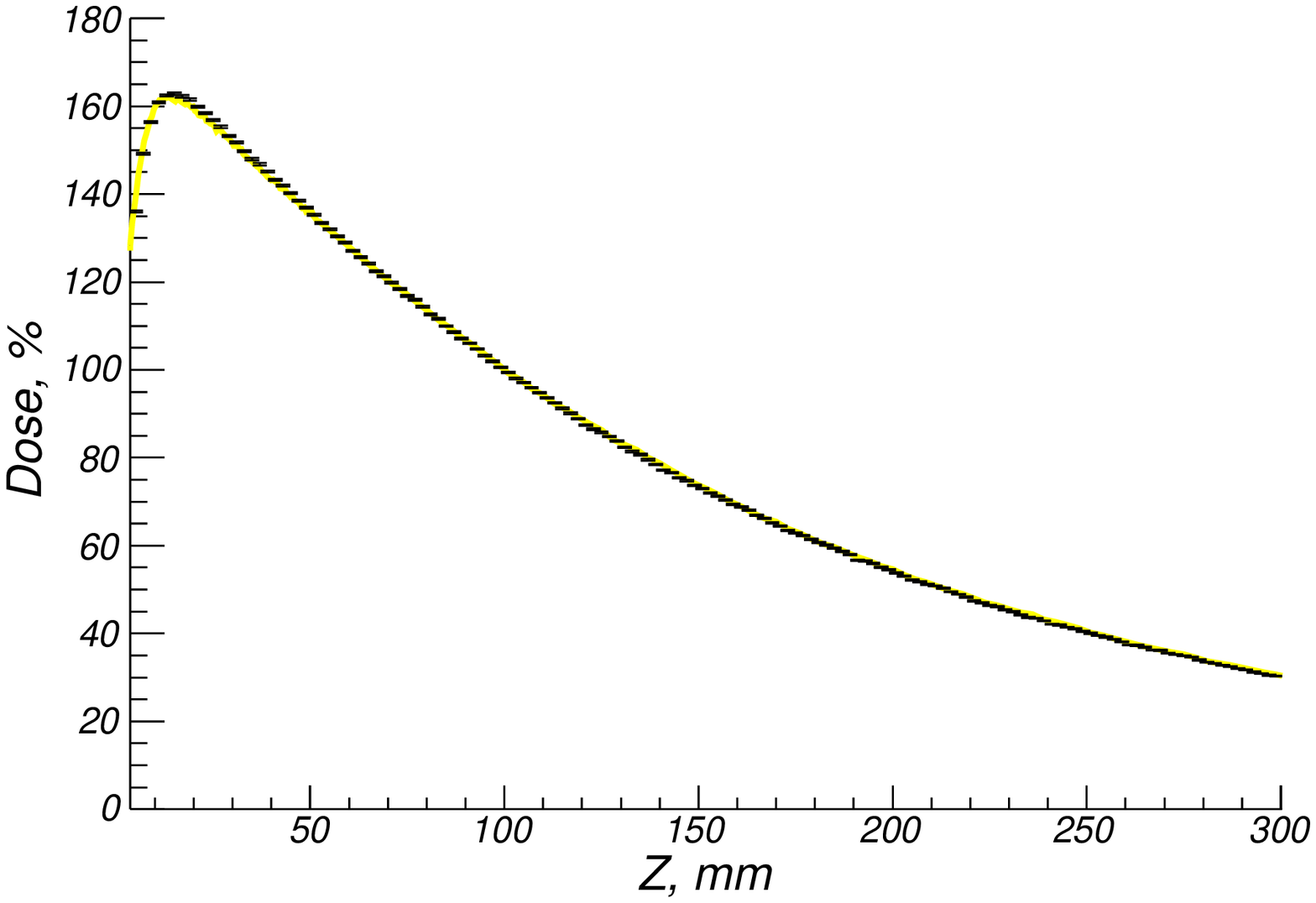}}
\subfigure[]{\includegraphics[totalheight=0.2\textwidth]{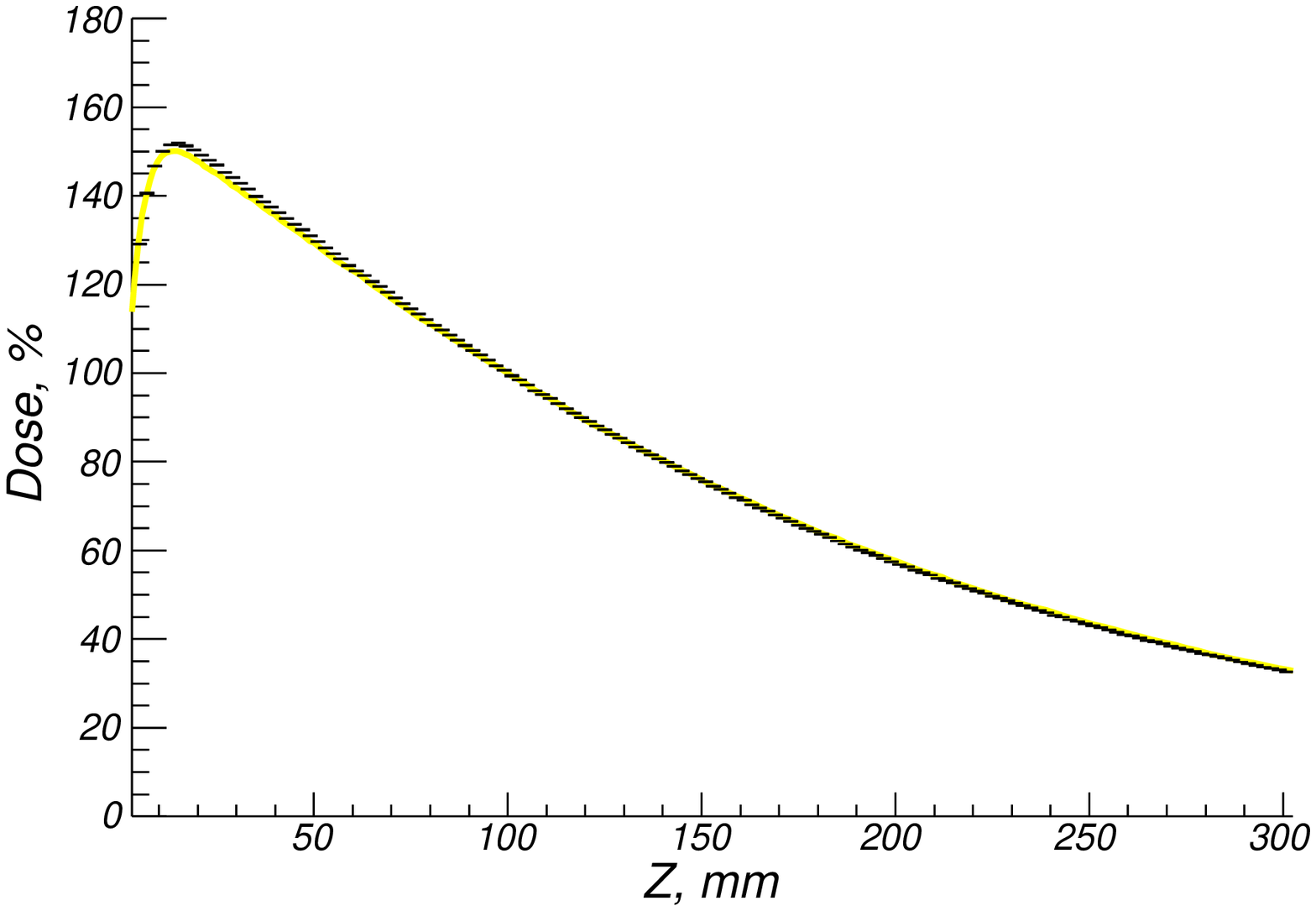}}
\caption{Simulated (data points) and experimental (solid line) PDDs for water phantom subject to irradiation by square fields of dimensions:  (a) $1\times1~cm^2$, (b) $4\times4~cm^2$, (c) $10\times10~cm^2$.}\label{fig:commissionPDD}
\end{figure}

Primary electron beam parameters that provided the best-fit to experimental PDDs and crossplane profiles at a depth of 15 mm in water phantom were found to be $E=6.0~MeV$ and $\sigma=1.1~mm$. This result is in contrast to published values for tuned beam energy that can often vary from the nominal beam energy by up to 5\% (Mesbahi \etal 2006, Verhaegen and Seuntjens 2003). Particular attention was paid to the geometry and composition of the key components of the LINAC. In particular, the correct density of target and flattening filter is essential, an over-estimation by a few percent in flattening filter density leads to significant underestimation of beam ``horns'' in the cross-plane profile for larger fields. Furthermore, the simulated PDD was particularly sensitive to the range cuts employed in the simulation. Too high a cut value leads to an over-estimation of the average beam energy (beam is too hard) which in turn can lead to an underestimation of the peak dose, requiring an unusually low tuned beam energy to achieve agreement. For each simulation $5\times10^8$ primary electrons were used and simulation times were approximately 2 hrs on 100 cores (2.33GHz 64bit Intel Xeon processor cores) of a high performance computing facility.

\begin{figure}[here]
\centering
\subfigure[]{\includegraphics[totalheight=0.2\textwidth]{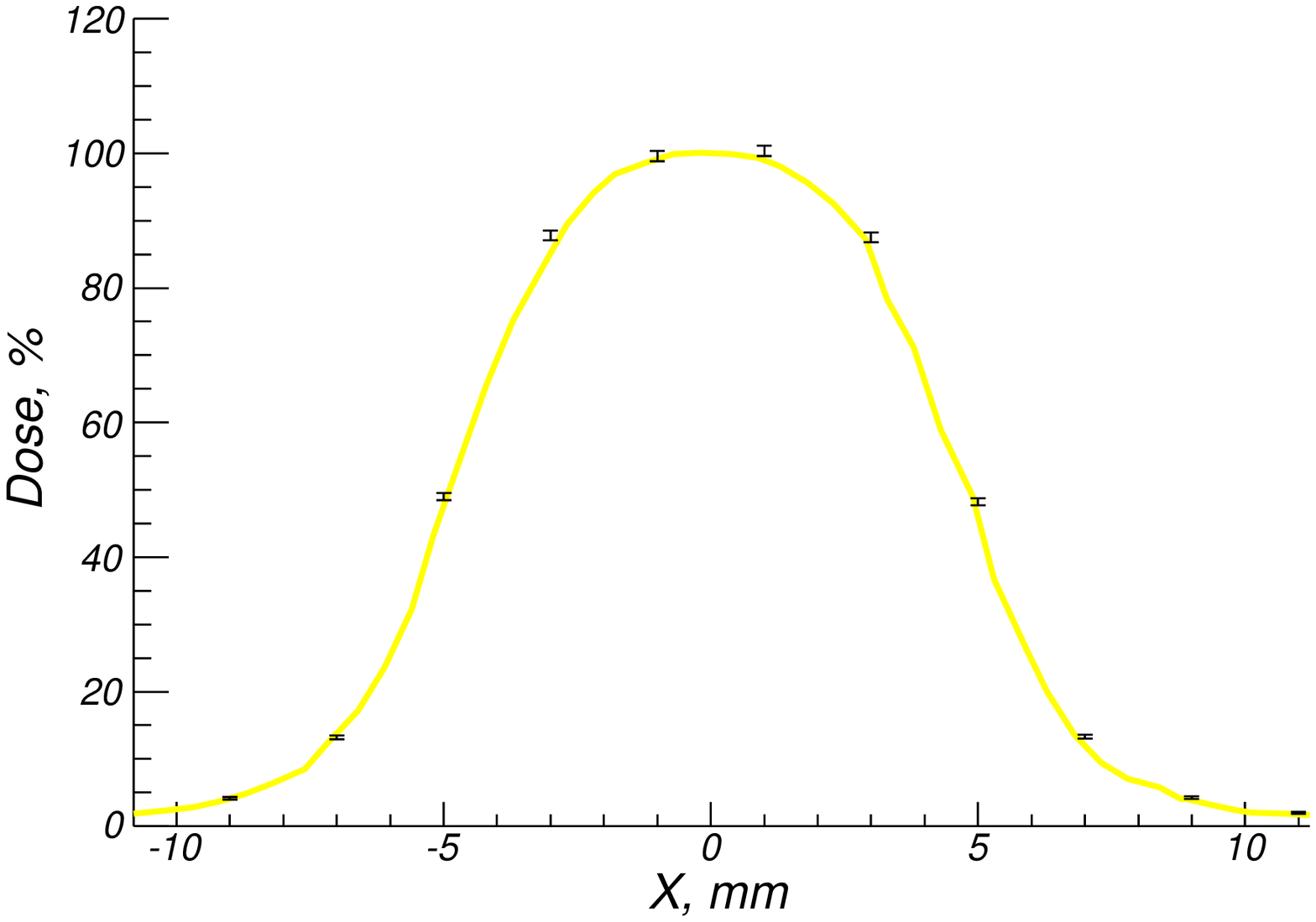}}
\subfigure[]{\includegraphics[totalheight=0.2\textwidth]{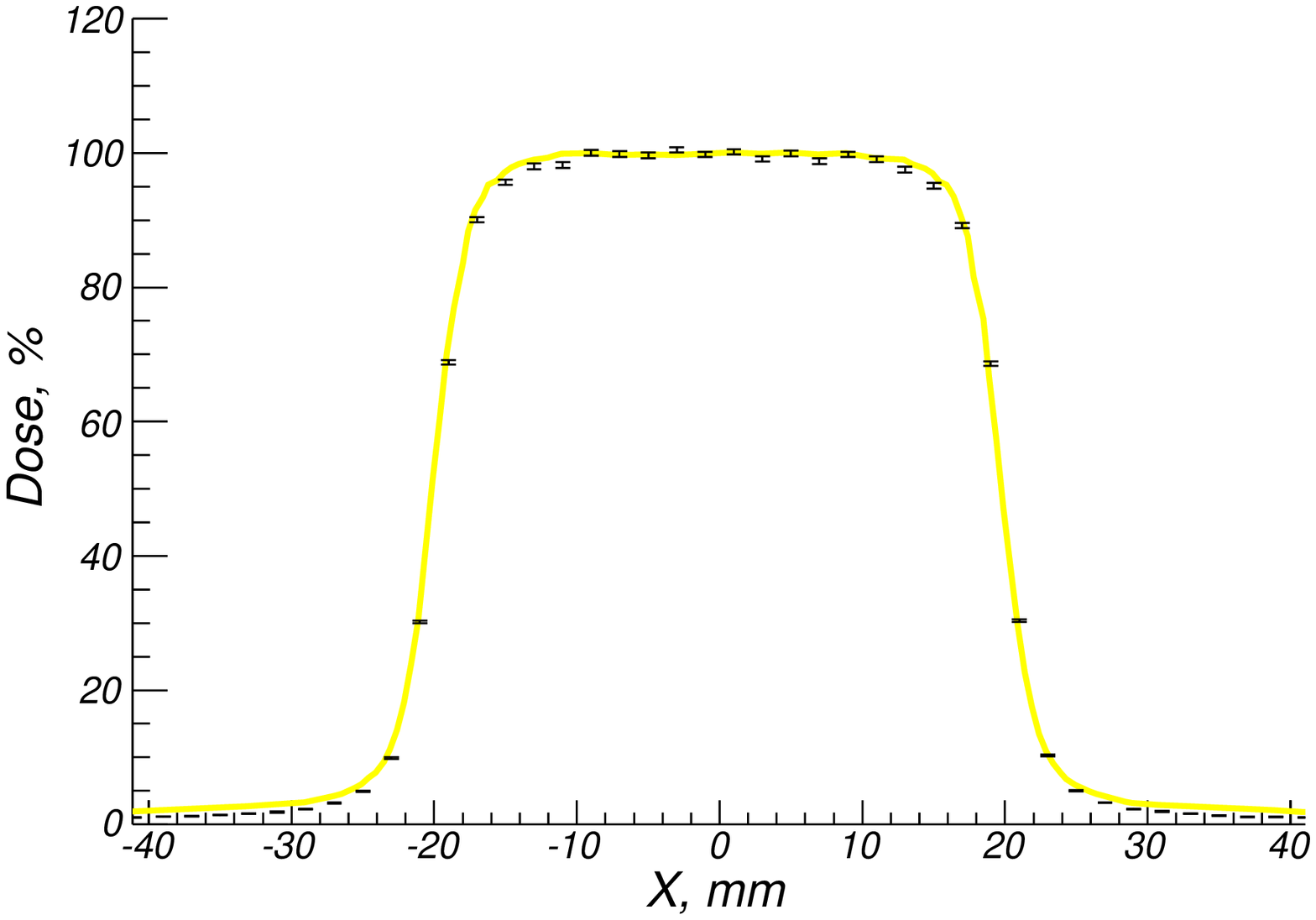}}
\subfigure[]{\includegraphics[totalheight=0.2\textwidth]{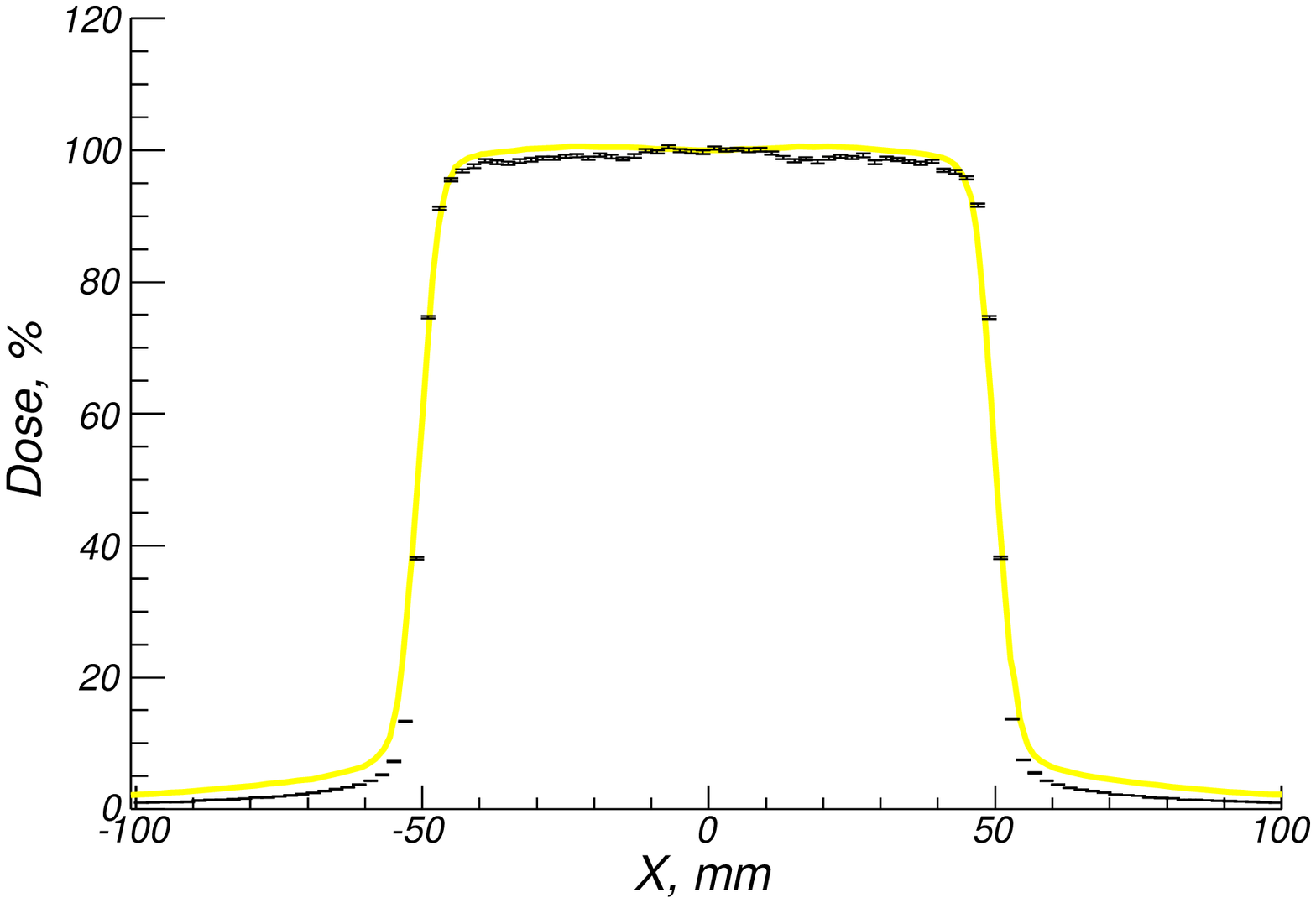}}
\subfigure[]{\includegraphics[totalheight=0.2\textwidth]{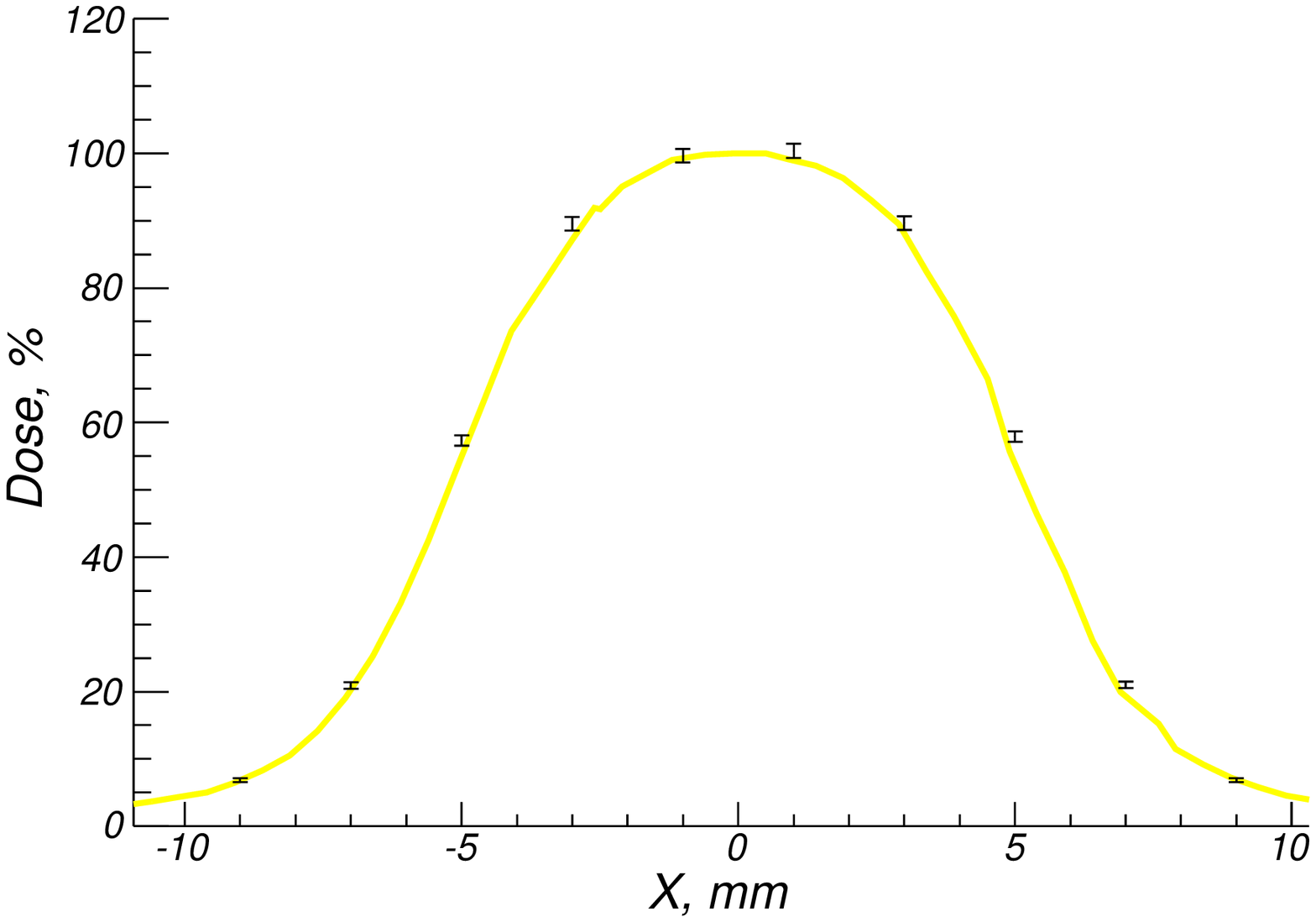}}
\subfigure[]{\includegraphics[totalheight=0.2\textwidth]{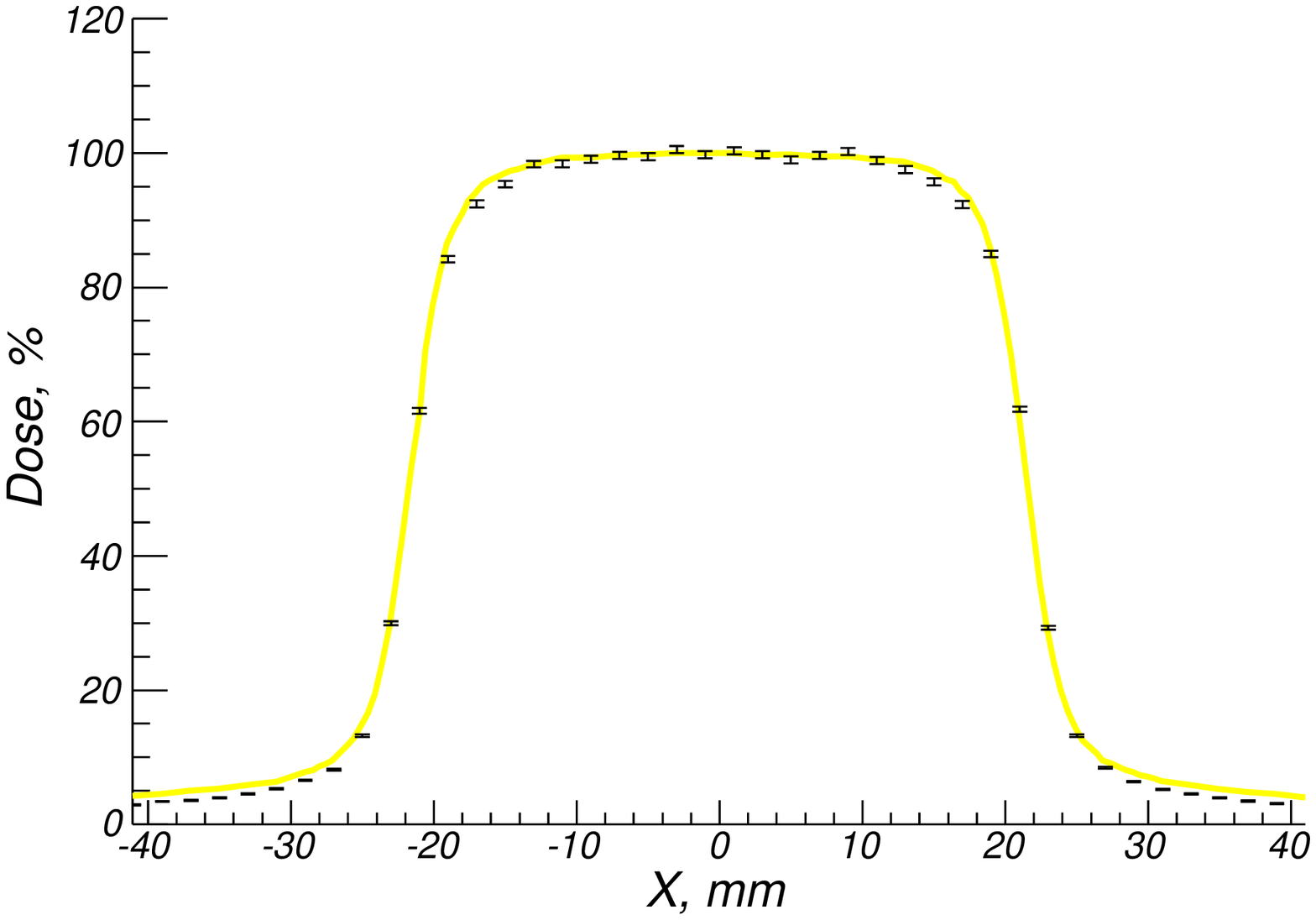}}
\subfigure[]{\includegraphics[totalheight=0.2\textwidth]{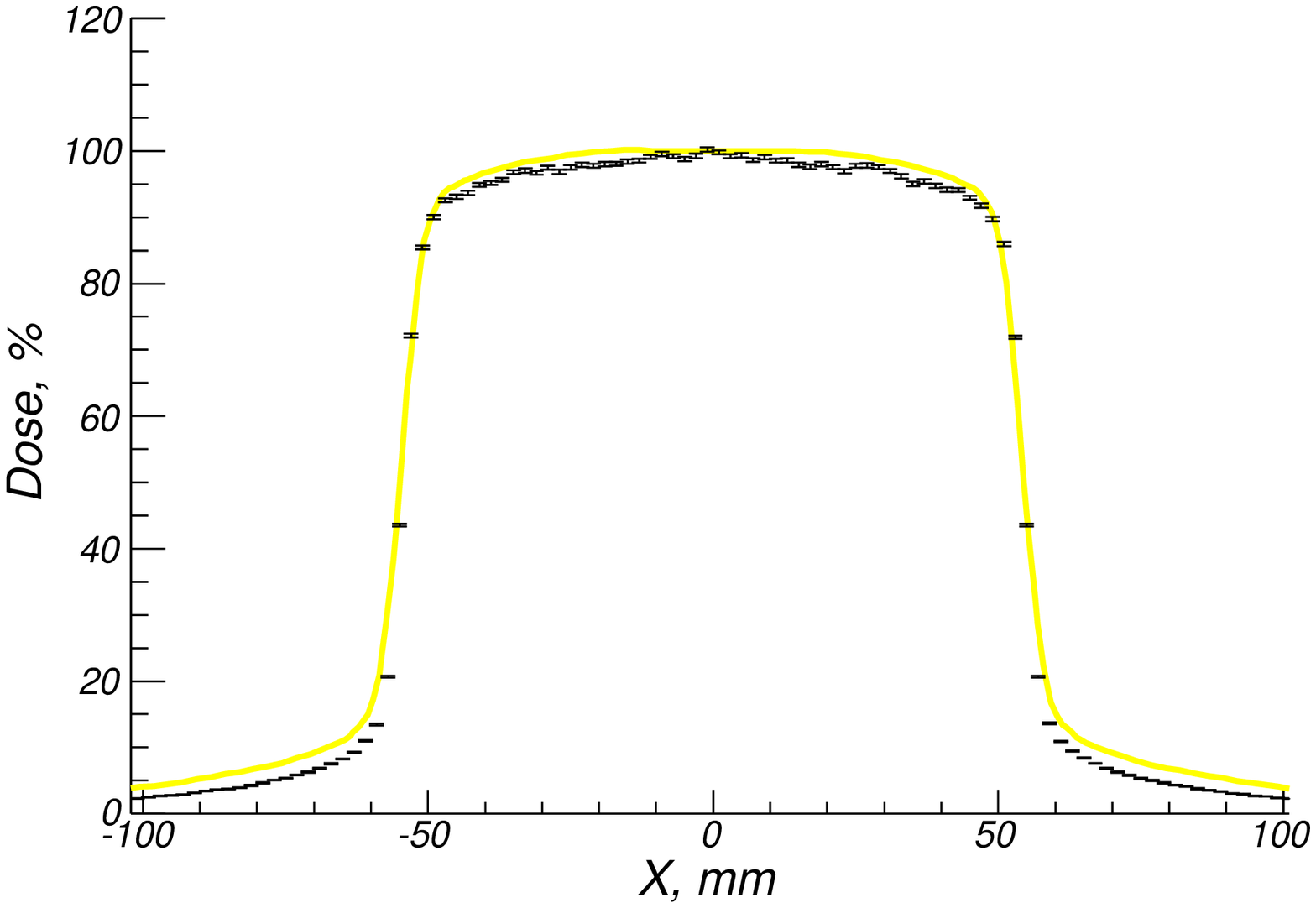}}
\subfigure[]{\includegraphics[totalheight=0.2\textwidth]{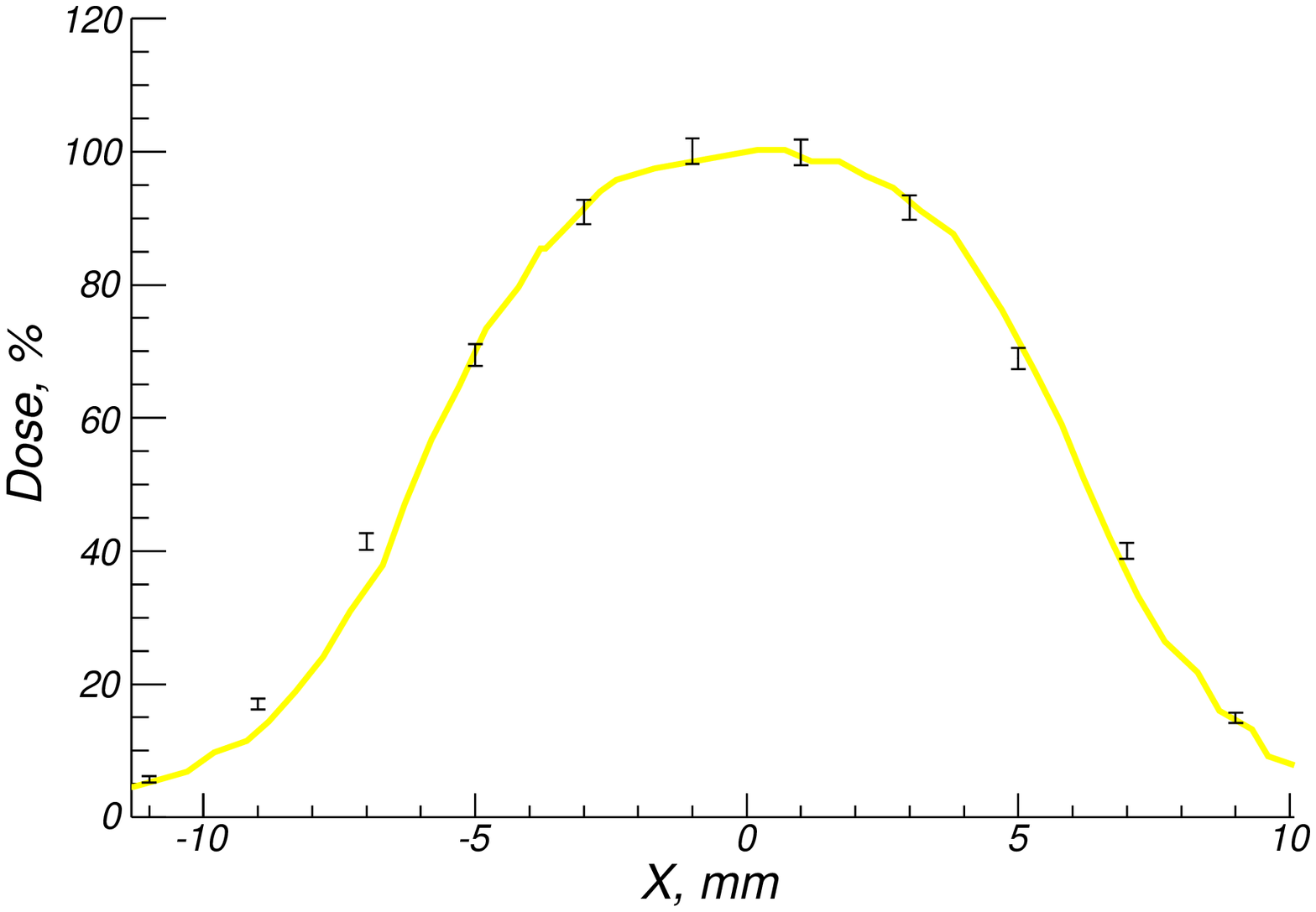}}
\subfigure[]{\includegraphics[totalheight=0.2\textwidth]{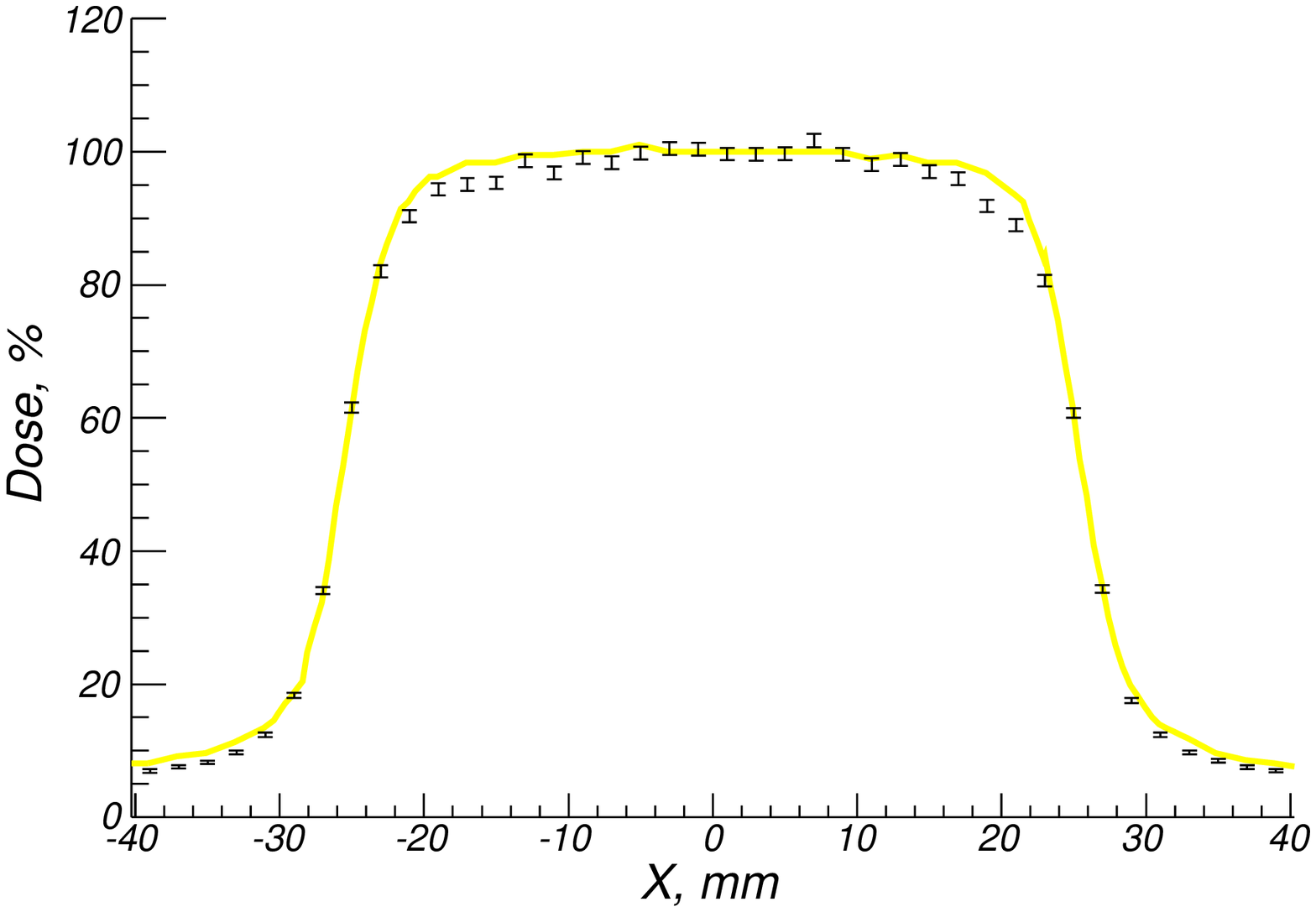}}
\subfigure[]{\includegraphics[totalheight=0.2\textwidth]{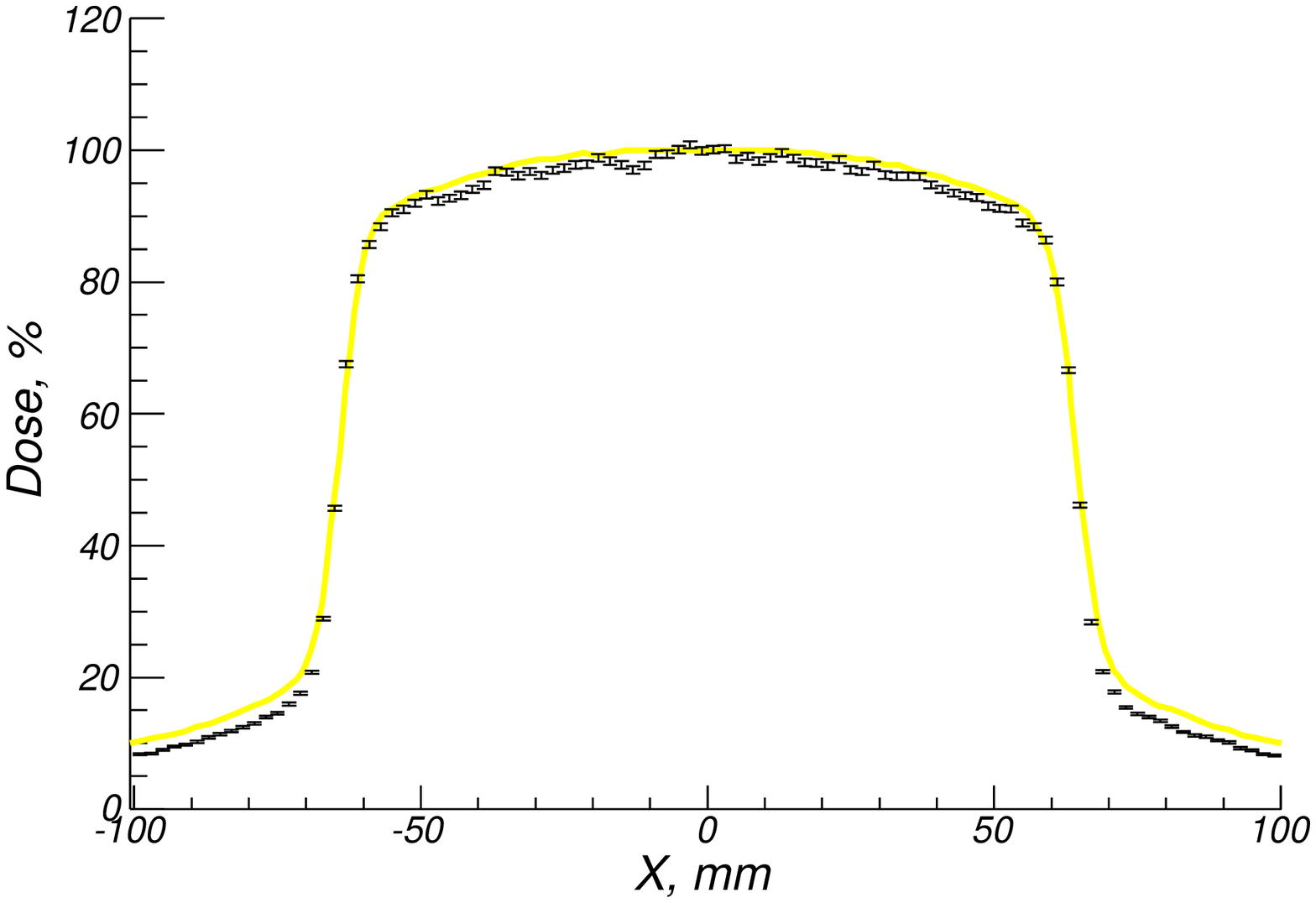}}
\caption{Simulated (data points) and experimental (solid line) crossplane dose profiles in water phantom for a 6MV tuned beam at depths of: (a,b,c) $15~mm$, (d,e,f) $100~mm$, and (g,h,i) $300~mm$ for field sizes of: (a,d,g) $1\times1$ cm$^2$, (b,e,h) $4\times4$ cm$^2$, (c,f,i) $10\times10$ cm$^2$}\label{fig:commissionProfile}
\end{figure}

Figure~\ref{fig:commissionPDD} shows PDD curves, both experimental and simulated, for a number of field sizes relevant to RapidArc and IMRT treatments. The curves are normalised to the dose at a depth of 10 cm. The uncertainty of simulation results is approximately 2\% and curves agree within the specified gamma criteria for at least 98\% of points.

Figure~\ref{fig:commissionProfile} shows crossplane dose profiles, both experimental and simulated, for a number of field sizes and depths in a water phantom. Agreement between experiment and simulation is observed to satisfy a gamma criterion of $3\%/3\,mm$ (98\% of data points) and verifies the accuracy of the primary beam model, geometry of the beam modifying components (barring the MLCs), physics processes, and the spectral properties of the 6 MV photon beam. 

\begin{figure}[here]
\centering
\subfigure[]{\includegraphics[height=0.2\textheight]{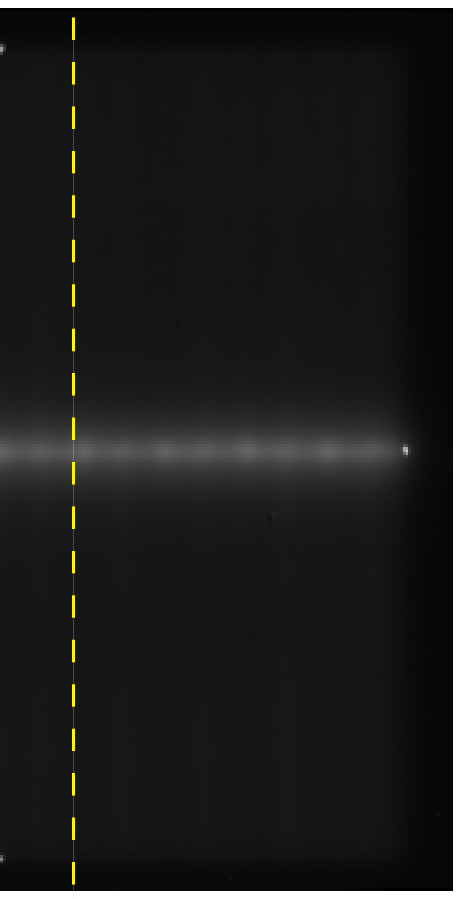}}
\subfigure[]{\includegraphics[width=0.2\textheight]{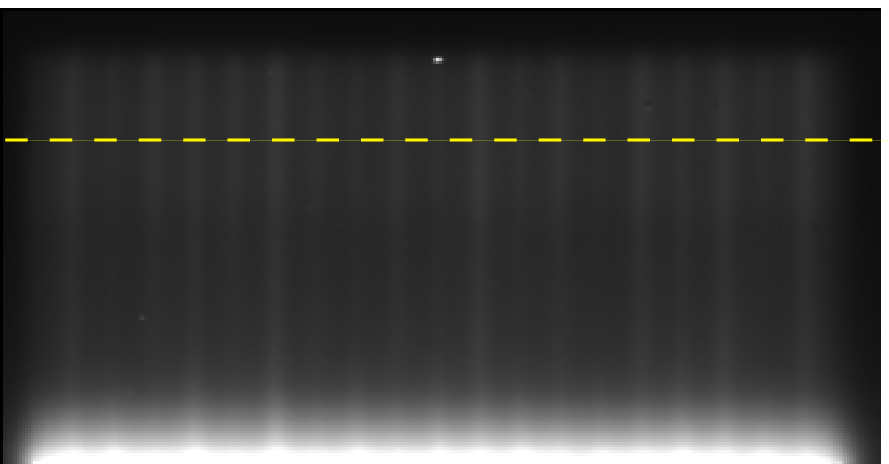}}
\caption{EBT2 film measurements of the 2D dose distributions for a blocked field: (a) low MU irradiation to determine abutted leaf leakage doses along with vertical profile used for comparison with simulation results at offset of $x=1\,cm$ from the y-axis at the left of image and (b) high MU irradiation to determine interleaf leakage doses, showing the x-profile used for comparison with simulation at offset of $y=4\,cm$ from x-axis along bottom of image. }
\label{fig:films}
\end{figure}

\begin{figure}[here]
\centering
\subfigure[]{\includegraphics[width=0.2\textheight]{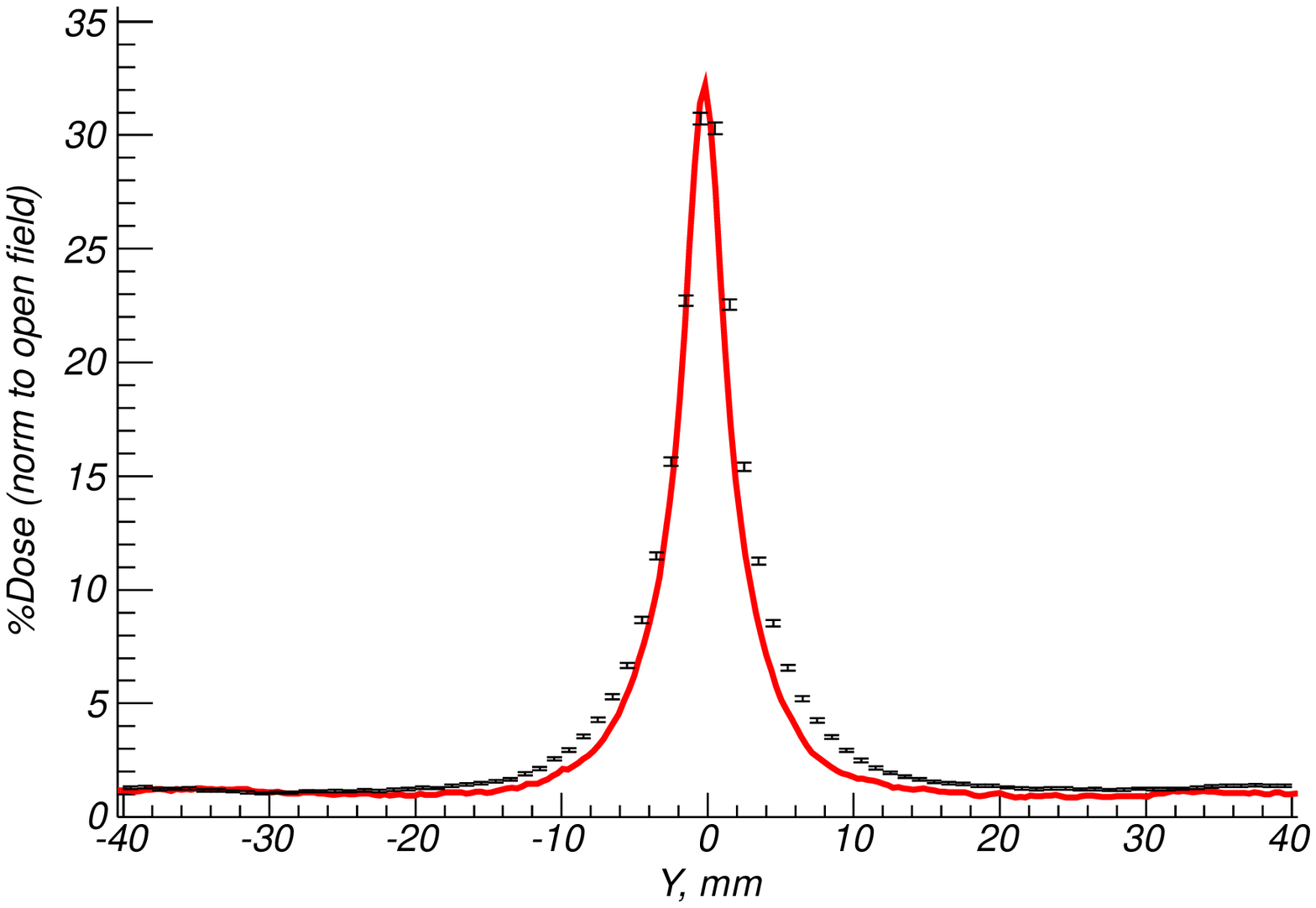}}
\subfigure[]{\includegraphics[width=0.2\textheight]{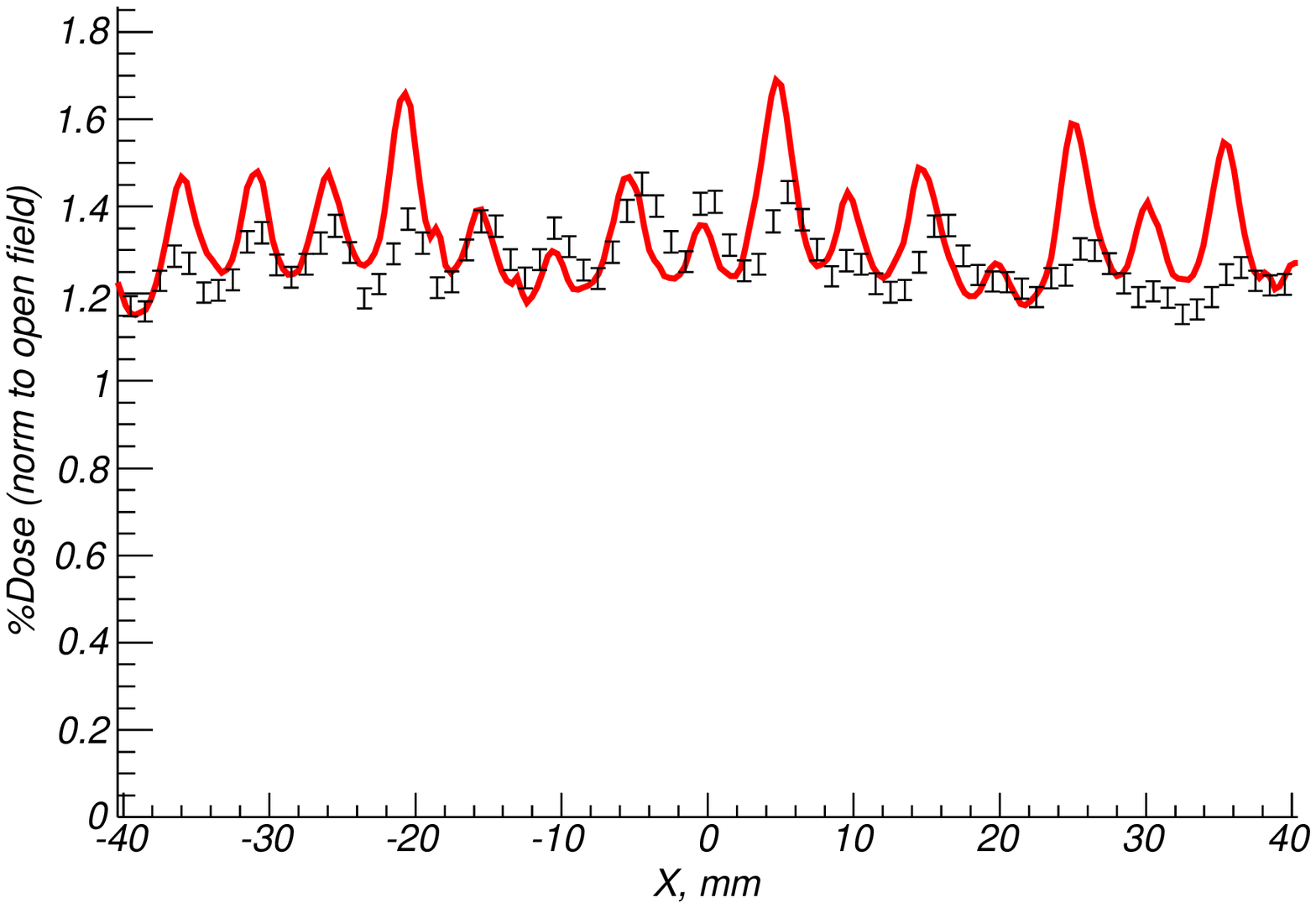}}
\caption{Simulated (data points) and experimental (solid line) profiles for (a) abutted leaf leakage and (b) interleaf leakage. Doses are normalised to open field dose for a $10\times10$cm$^2$ field.}
\label{fig:leakage}
\end{figure}

In order to mitigate the problem of inhomogeneity of EBT2 film sensitivity reported in the literature (Kairn \etal 2010), a simple correction technique was used for which two pieces from the same sheet were used to measure the dose distribution with one of the pieces rotated 180 degrees. Using fiducial markers, the resulting dose distributions were co-registered and averaged. Two leakage profiles were considered for comparison with simulation predictions (see figure~\ref{fig:films}): the first being parallel to the direction of leaf travel on the beam axis, referred to herein as the abutted leaf leakage profile; the second was perpendicular to the direction of leaf travel and offset from the beam axis by several centimetres, referred to here-in as the interleaf leakage profile. The results of the abutted and interleaf leakage profiles are shown in figure~\ref{fig:leakage}, dose profiles are normalised to the dose at the same position in the phantom for a $10\times10\,$cm$^2$ square field. The abutted leaf leakage profiles show excellent agreement, satisfying the criterion of $3\%/3\,mm$. The interleaf leakage profiles show good agreement between the mean value of leakage dose: $1.33\pm0.11\%$ (standard deviation) measured compared to $1.27\pm0.07\%$ simulated. Both in good agreement with the results of Heath and Seuntjens (2003) for Millenium MLC modelling using BEAMnrc. Heath and Seuntjens (2003) observed a discrepancy at $y >10mm$ for the abutted leaf leakage profiles and attributed this to the existence of a single calibration point below $20\,cGy$ on the calibration curve. This effect was also observed during the current study and mitigated by additional low dose calibration points below $20\,cGy$.

\begin{figure}[here]
\centering
\subfigure[]{\includegraphics[width=0.3\textwidth]{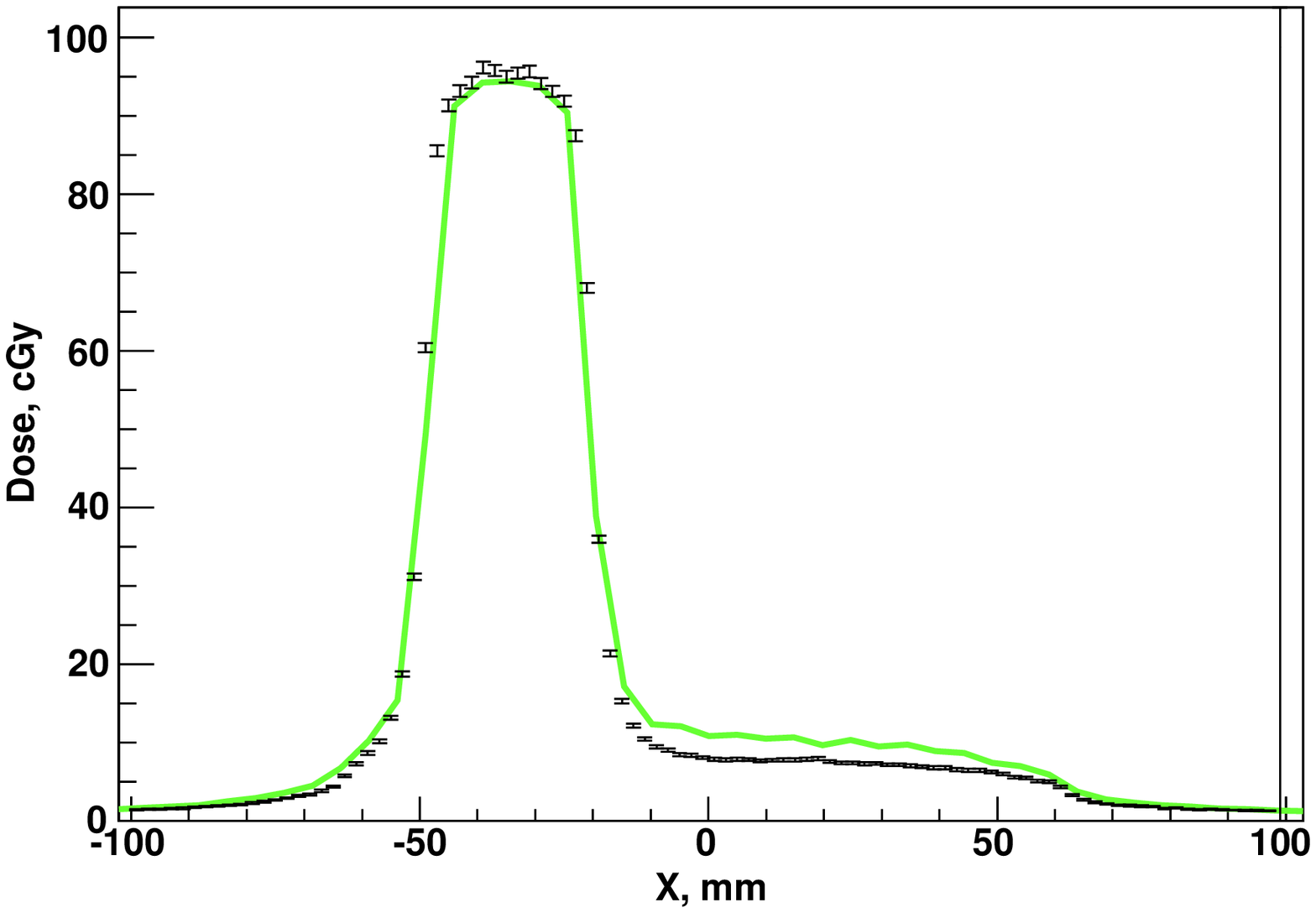}}
\subfigure[]{\includegraphics[width=0.3\textwidth]{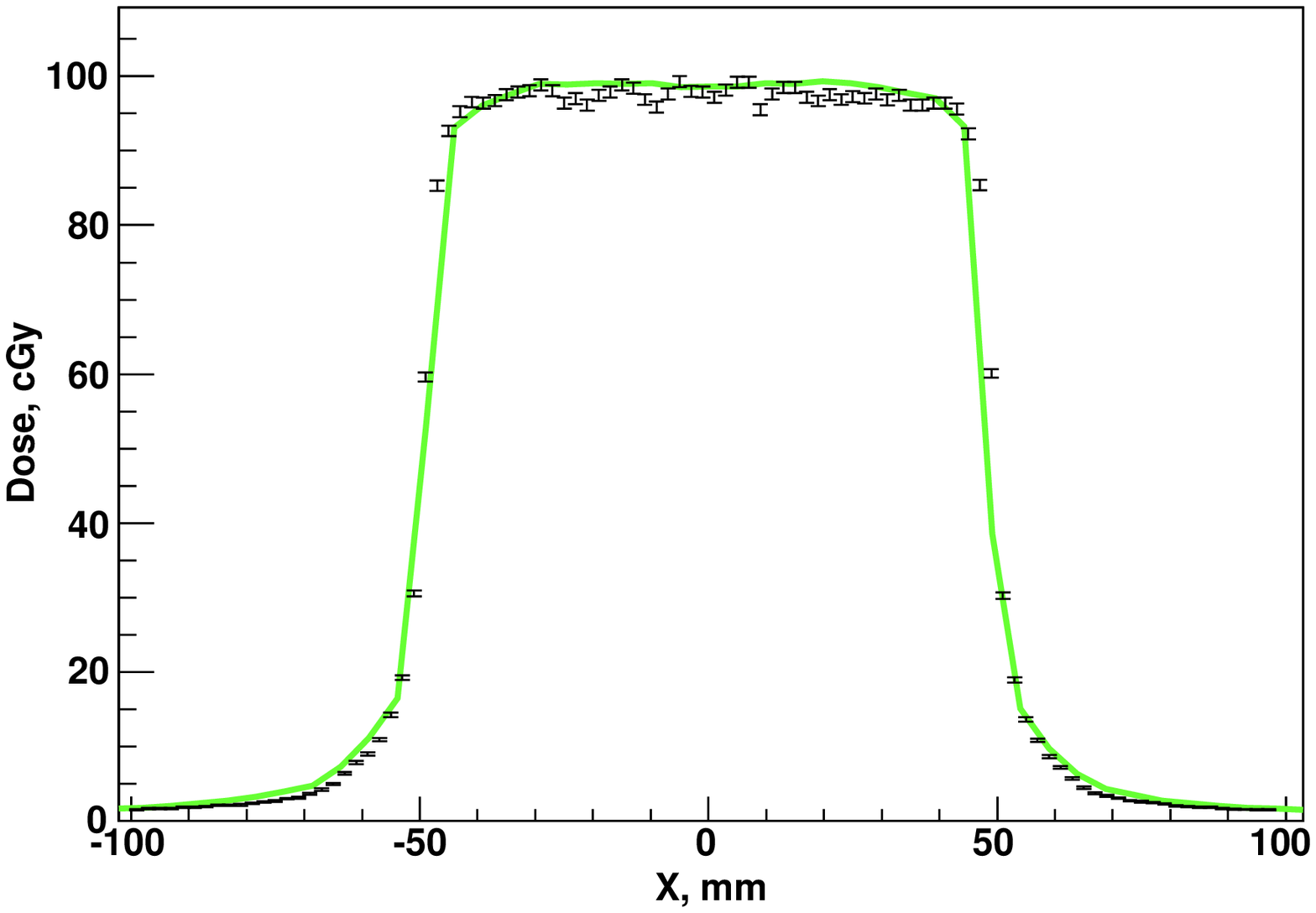}}
\subfigure[]{\includegraphics[width=0.3\textwidth]{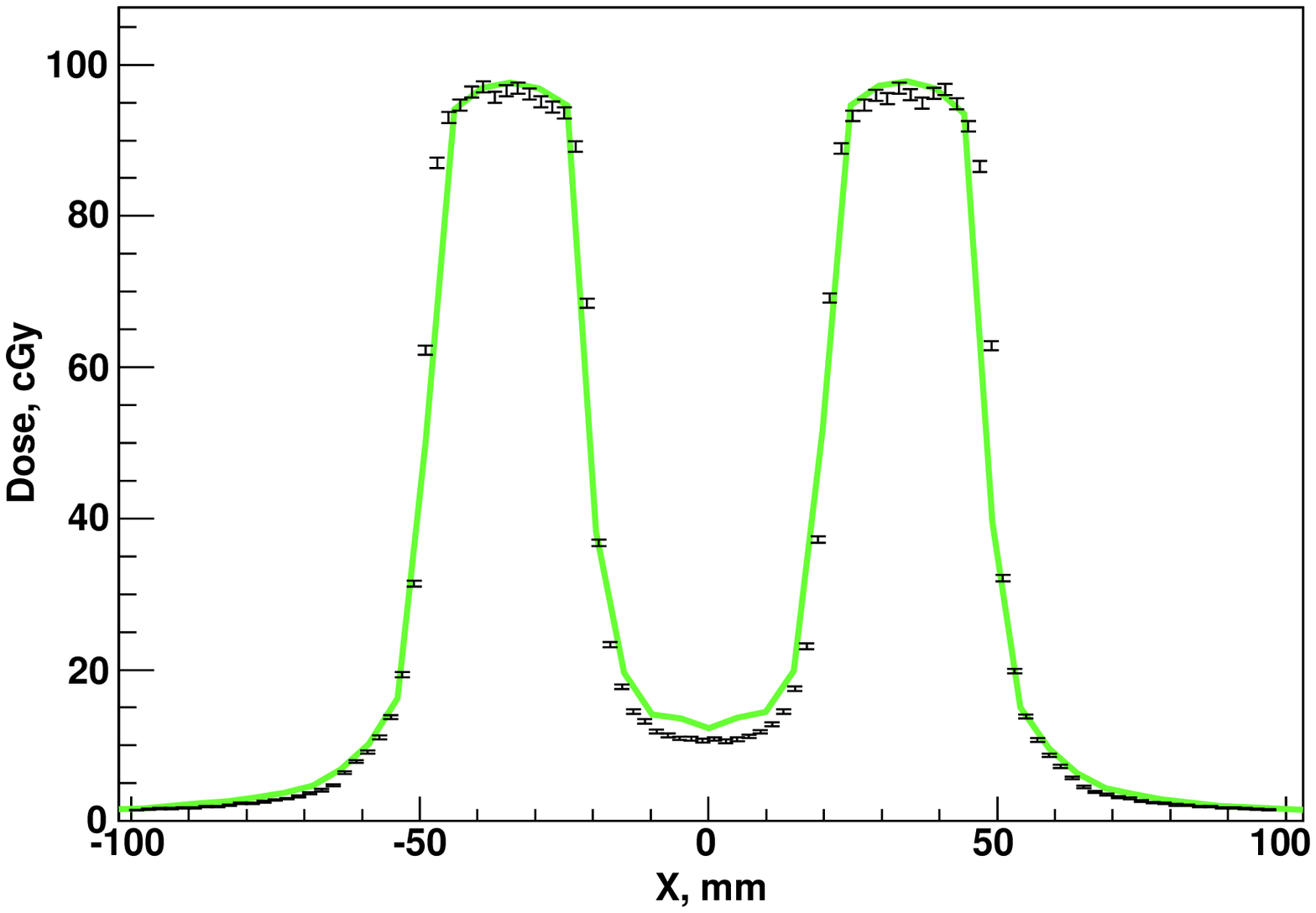}}
\caption{Simulated (data points) and experimental (solid line) profiles as measured by the MapCHECK array exposed to the chair-test treatment plan. X-profiles are at y offsets of (a) $+40\,mm$, (b) $0\,mm$, and (c) $-40\,mm$.}
\label{fig:chair}
\end{figure}

Results of the chair-test measurement and simulation are shown in figure~\ref{fig:chair}.
Figure~\ref{fig:chair}(b) shows an x-profile along the y-axis, through the `seat' of the chair, where agreement between simulation and experiment is good. Figure~\ref{fig:chair}(a) shows an x-profile at a y-offset of $+40\,mm$; ie., across the back of the chair. Discrepancy in absolute doses of around 4\% between experiment and simulation can be seen for the region that is blocked by the MLCs for the entire irradiation ($X>-10\,mm$). The discrepancy was thought to be due to an over-response by the MapCHECK array to the lower energy photons in this region (dominated by particles scattered in the phantom or transmitted through the leaves). To test this hypothesis, leaf leakage measurements shown in Figure~\ref{fig:leakage} were repeated using the MapCHECK device. This yielded an interleaf leakage measurement of $1.38\pm0.106\%$, around $8\%$ higher than simulation and film measurements, yet agreeing within the stated uncertainty. Figure~\ref{fig:chair}(c) shows an x-profile at a y-offset of $+40\,mm$; ie., across the legs of the chair. Again discrepancy is seen in the central region that is either directly under a stationary leaf-body, or traversed by fully closed leaves. Despite these discrepancies, these results do confirm the ability of the tool to simulate the control points of a treatment plan, accurately reproducing leaf movements, number of MUs per control point, as well as the method of absolute dose calculation.

\section{Conclusions}
A Geant4 based simulation tool has been developed that is capable of accurately simulating the dosimetric properties of a 6 MV Varian$\textsuperscript{\textregistered}$ iX clinac. The simulation includes detailed modelling of key components using CAD software and was tuned and validated against both film and ionisation chamber dosimetry measurements in solid water and water phantoms. The accuracy of the multi-leaf collimator model and DICOM-RT interface was verified against MapCHECK measurements in a solid water phantom subject to irradiation by a chair-test. This tool will form the basis of a treatment plan verification tool for radiotherapy and the model will be extended to include higher energy beams as well as CAD modelled electron applicators for modelling electron beams. Further validation will be performed by simulating RapidArc treatment plans delivered to homogeneous solid water phantoms and anthropomorphic phantoms and comparisons with experimental measurement.

\section*{Acknowledgements}
This project is funded by the Queensland Cancer Physics Collaborative (Queensland Health), Australia. Computational resources and services used in this work were provided by the HPC and Research Support Unit, Queensland University of Technology, Brisbane, Australia (404 core Lyra Altix SE compute cluster). The authors would like to thank the Geant4 collaboration for providing the toolkit and examples, regular updates, documentation, and the online user forum. The authors would also like to thank Tanya Kairn and John Kenny of Premion for discussions and guidance related to the use of EBT2 film. 

\section*{References}

\begin{harvard}{}

\item[] Allison J {\it et al} 2006 Geant4 developments and applications {\it IEEE Trans. Nucl. Sci.}  {\bf 53} 270--278

\item[] Aso T, Kimura A, Yamashita T and Sasaki T 2007 Optimization of patient geometry based on CT data in GEANT4 for medical application {\it IEEE Nucl. Sci. Symp. Conf. Rec.} {\bf 4} 2576--2580

\item[] Bush K, Townson R and Zavgorodni S 2008 Monte Carlo simulation of RapidArc radiotherapy delivery \PMB {\bf 53} N359--N370


\item[] Constantin M, Constantin D E, Keall P J, Narula A, Svatos M and Perl J 2010 Linking computer-aided design (CAD) to Geant4-based Monte Carlo simulations for precise implementation of complex treatment head geometries \PMB {\bf 55} N211--N220 

\item[] Dong X, Cooperman G and Apostolakis J 2010 Multithreaded Geant4: Semi-automatic Transformation into Scalable Thread-Parallel Software {\it Lect. in Comp. Sci} {\bf 6272} 287--303

\item[] Faddegon B A, Asai M, Perl J, Ross C, Sempau J, Tinslay J, Salvat F 2008 Benchmarking of Monte Carlo simulation of bremsstrahlung from thick targets at radiotherapy energies {\it Med. Phys.} {\bf 35(10)} 4308--4317

\item[] Foppiano F, Mascialino B, Pia M.G and Piergentili M 2004 A Geant4-based simulation of an accelerator's head used for intensity modulated radiation therapy {\it IEEE Nucl. Sci. Symp. Conf. Rec.} {\bf 4} 2128--2132

\item[] Habib B, Poumarede B, Tola F and Barthe J 2009 Evaluation of PENFAST A fast Monte Carlo code for dose calculations in photon and electron radiotherapy treatment planning {\it Phys. Medica} {\bf 26} 17--25

\item[] Hartmann B, Martisikova M and Jakel O 2010 Technical Note: Homogeneity of Gafchromic EBT2 film {\it Med. Phys.} {\bf 37} 1753--1757

\item[] Heath E and Seuntjens J 2003 Development and validation of a BEAMnrc component module for accurate Monte Carlo modelling of the Varian dynamic Millennium multileaf collimator \PMB {\bf 48} 4045--4064

\item[] International Atomic Energy Agency 2000 Absorbed Dose Determination in 
External Beam Radiotherapy {\it IAEA Technical Report Series no 398}

\item[] International Electrotechnical Commission 2002 Radiotherapy equipment – Coordinates, movements and scales {\it Technical Report 61217}

\item[] Jan S {\it et al} 2011 GATE V6: a major enhancement of the GATE simulation platform enabling modelling of CT and radiotherapy \PMB {\bf 56} 881--901

\item[] Grevillot L, Frisson T, Maneval D, Zahra N, Badel J-N and Sarrut D 2011 Simulation of a 6 MV Elekta Precise Linac photon beam using GATE/GEANT \PMB {\bf 56} 903--918

\item[] Kairn T, Aland T and Kenny J 2010 Local heterogeneities in early batches of EBT2 film: a suggested solution \PMB {\bf 55} L37-42

\item[] Kawrakow I and Walters B R B 2006 Efficient photon beam dose calculations using DOSXYZnrc with BEAMnrc {\it Med. Phys.} {\bf 33} 3046--3057

\item[] Li X A, Ma L, Naqvi S, Shih R and Yu C 2001 Monte Carlo dose verification for intensity-modulated arc therapy \PMB {\bf 46} 2269--2283

\item[] Mesbahi A, Reilly A J, Thwaites D I 2006 Development and commissioning of a Monte Carlo photon beam model for Varian Clinac 2100EX linear accelerato {\it Appl. Radiat. Isot.} {\bf 64} 656--662

\item[] Othman M A R, Cutajar D L, Hardcastle N, Guatelli S and Rosenfeld A B 2010 Monte carlo study of MOSFET packaging, optimised for improved energy response: single MOSFET filtration {\it Rad. Prot. Dos.} {\bf 141(1)} 10--17

\item[] Locke C and Zavgorodni S 2008 Vega library for processing DICOM data required in Monte Carlo verification of radiotherapy treatment plans {\it Australas. Phys. Eng. Sci. Med.} {\bf 31} 290--299

\item[] Low D A, Harms W B, Mutic S and Purdy J A 1998 A technique for the quantitative evaluation of dose distributions {\it Med. Phys.} {\bf 25} 656--661

\item[] Oborn B M, Metcalfe P E, Butson M J and Rosenfeld A B 2009 High resolution entry and exit Monte Carlo dose calculations from a linear accelerator 6 MV beam under the influence of transverse magnetic fields {\PMB} {\bf 36(8)} 3549--3559

\item[] Pisaturo O, Moeckli R, Mirimanoff R-O and Bochud F O 2009 A Monte Carlo-based procedure for independent monitor unit calculation in IMRT treatment plans {\it Phys. Med. Biol.} {\bf 54} 4299--4310

\item[] Paganetti H 2004 Four-dimensional Monte Carlo simulation of time-dependent geometries {\PMB} {\bf 49(6)} N75--N81
 
\item[] Poon E and Verhaegen F 2005 Accuracy of the photon and electron physics in GEANT4 for radiotherapy applications {\it Med. Phys.} {\bf 32} 1696-1711
 
\item[] Popescu I A, Shaw C P, Zavgorodni S F and W A Beckham 2005 Absolute dose calculations for Monte Carlo simulations of radiotherapy beams \PMB {\bf 50} 3375--3392

\item[] Spezi E, Lewis D G and Smith C W 2002 A DICOM-RT-based toolbox for the evaluation and verification of radiotherapy plans \PMB {\bf 47} 4223--4232

\item[] Tinslay J, Perl J and Asai M 2007 Verification of Bremsstrahlung Splitting in Geant4 for Radiotherapy Quality Beams {\it Med. Phys.} {\bf 34(6)} 2504--2510

\item[] Van Esch A, Bohsung J, Sorvari P, Tenhunen M, Paiusco M, Iori M, Engstr\"{o}m P, Nystr\"{o}m H and Huyskens D P 2002 Acceptance tests and quality control (QC) procedures for the clinical implementation of intensity modulated radiotherapy (IMRT) using inverse planning and the sliding window technique: experience from five radiotherapy departments. {\it Radiother. Oncol.} {\bf 65(1)} 53--70 

\item[] Verhaegen F and Seuntjens J 2003 Monte Carlo modelling of external radiotherapy photon beams \PMB {\bf 48} R107--164

\item[] Wieslander E and Kn\"{o}\"{o}s T 2000 A virtual linear accelerator for verification of treatment planning systems \PMB {\bf45(10)} 2887--2896

\item[] Wroe A, Rosenfeld A and Schulte R 2007 Out-of-field dose equivalents delivered by proton therapy of prostate cancer {\it Med. Phys.} {\bf 34} 3449--3457

\end{harvard}

\end{document}